\documentclass[prb,twocolumn,superscriptaddress]{revtex4-2}
\usepackage{amsfonts}
\usepackage{amssymb}
\usepackage{amsmath}
\usepackage{graphicx}
\usepackage{dcolumn}
\usepackage{bm}
\usepackage{units}
\usepackage{soul}
\usepackage{multirow}
\usepackage{mathrsfs} 
\usepackage{CJKutf8}
\usepackage{subfigure}
\usepackage{color}%
\usepackage{url}
\usepackage[colorlinks,linkcolor=blue,anchorcolor=blue,
citecolor=blue,urlcolor=blue]{hyperref}
\usepackage{array}
\usepackage{ulem}
\begin{document}
\title{On universal butterfly and antisymmetric magnetoresistances}

\author{H. T. Wu}
\affiliation{Department of Physics, The Hong Kong University of Science and Technology, 
Clear Water Bay, Kowloon, Hong Kong, China}
\affiliation{HKUST Shenzhen Research Institute, Shenzhen, 518057, China}

\author{Tai Min}
\affiliation{Center for Spintronics and Quantum Systems, State Key Laboratory for Mechanical Behavior 
of Materials, Xi'an Jiaotong University, No.28 Xianning West Road, Xi'an, Shaanxi, 710049, China} 

\author{Z. X. Guo}
\affiliation{Center for Spintronics and Quantum Systems, State Key Laboratory for Mechanical Behavior 
of Materials, Xi'an Jiaotong University, No.28 Xianning West Road, Xi'an, Shaanxi, 710049, China} 

\author{X. R. Wang}
\email{phxwan@ust.hk}
\affiliation{Department of Physics, The Hong Kong University of Science and Technology, 
Clear Water Bay, Kowloon, Hong Kong, China}
\affiliation{HKUST Shenzhen Research Institute, Shenzhen, 518057, China}

\begin{abstract}
Butterfly magnetoresistance (BMR) and antisymmetric magnetoresistance (ASMR) 
are about a butterfly-cross curve and a curve with one peak and one valley
when a magnetic field is swept up and down along a fixed direction. 
Other than the parallelogram-shaped magnetoresistance-curve (MR-curve) 
often observed in magnetic memory devices, BMR and ASMR are two ubiquitous 
types of MR-curves observed in diversified magnetic systems, including van 
der Waals materials, strongly correlated systems, and traditional magnets. 
Here, we reveal the general principles and the picture behind the BMR and 
the ASMR that do not depend on the detailed mechanisms of magnetoresistance: 
1) The systems exhibit hysteresis loops, common for most magnetic materials 
with coercivities. 2) The magnetoresistance of the magnetic structures in 
a large positive magnetic field and in a large negative magnetic field
is approximately the same. With the generalized Ohm's law in magnetic 
materials, these principles explain why most BMR appears in the longitudinal 
resistance measurements and is very rare in the Hall resistance measurements. 
Simple toy models, in which the Landau-Lifshitz-Gilbert equation governs 
magnetization, are used to demonstrate the principles and explain the 
appearance and disappearance of BMR in various experiments. Our finding 
provides a simple picture to understand magnetoresistance-related experiments.
\end{abstract}

\maketitle

\section{Introduction}

\begin{figure}
\centering
\includegraphics[width=\columnwidth]{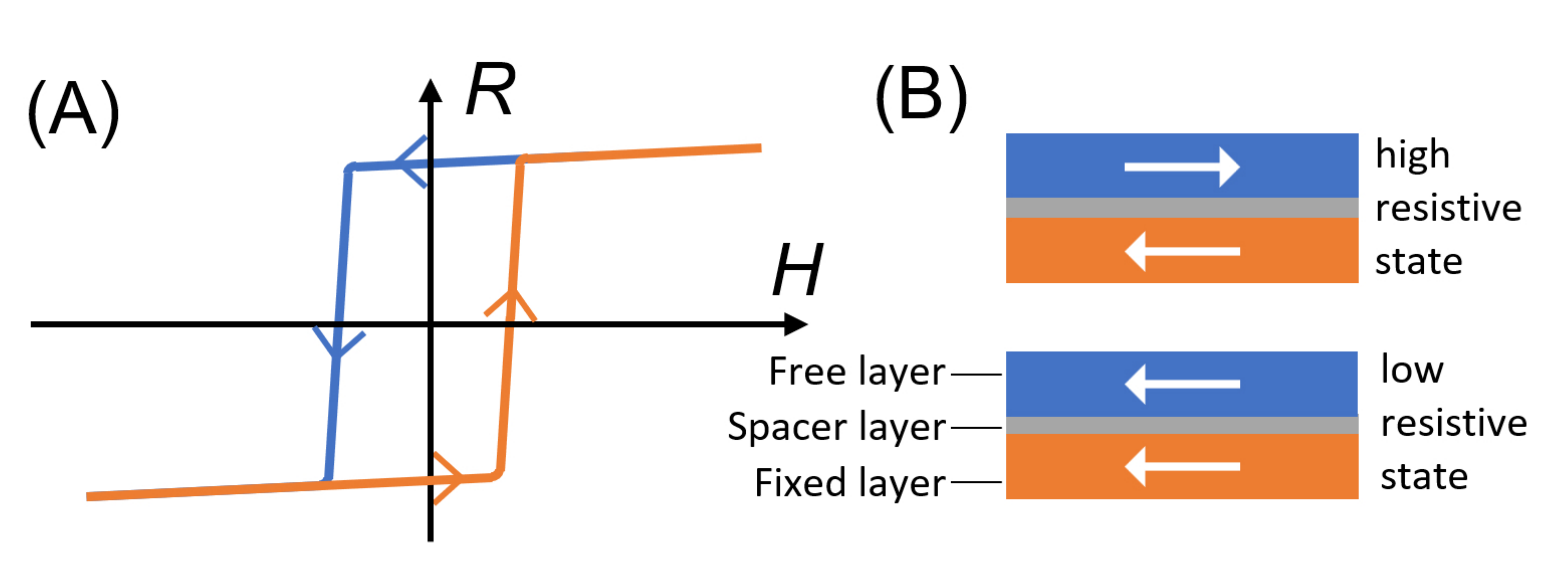}
\caption{(A) Schematics of parallelogram MR-curves when a 
device moves between a higher resistive state and a lower 
one in a field sweeping-up and sweeping-down process. 
(B) Illustration of a memory device with two stable 
resistive states. The device is in a lower (higher)
resistive state when the magnetization of the free-layer 
is parallel (antiparallel) with that of the fixed layer.}
\label{Fig1}
\end{figure}

\begin{figure*}
\centering
\includegraphics[width=2\columnwidth]{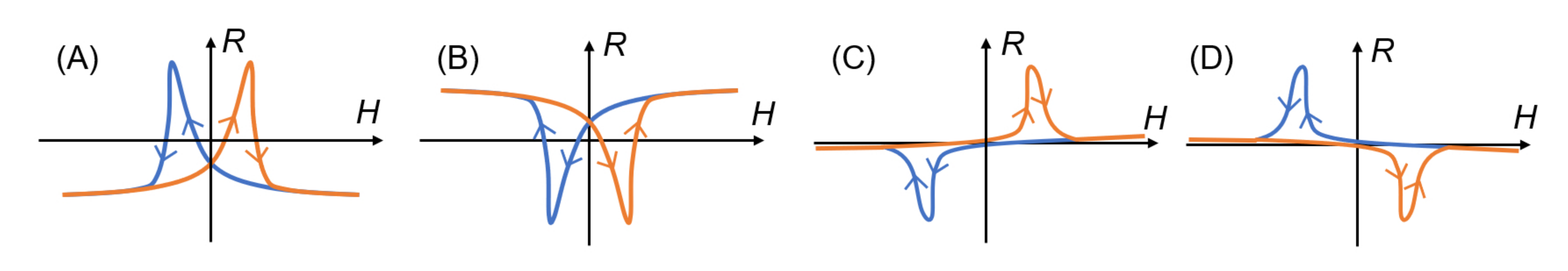}
\caption{Illustration of butterfly magnetoresistance 
(BMR) and antisymmetric magnetoresistance (ASMR) curves.
The orange and blue curves indicate resistance in the 
sweeping-up and sweeping-down processes, respectively.
(A)-(B) Upward (A) and downward (B) butterfly MR-curves that 
display a butterfly cross. (C)-(D) Two possible ASMR-curves 
that have the feature of one peak and one valley.}
\label{Fig2}
\end{figure*}

Magnetoresistance (MR) is an important quantity that is often used to 
probe and to understand the electronic properties of a condensed 
matter \cite{grosso2000}. Weak field MRs at low temperature are a   
standard probe for extracting quantum coherence length and time of 
metals \cite{Huckestein1995}, and high field MRs are a powerful 
tool for measuring the Fermi surfaces of metals \cite{grosso2000}. 
In magnetic materials with magnetic hysteresis, MR-curves can be 
classified into several types. One commonly-observed curve in 
magnetic memory devices is the parallelogram shape as shown in 
Fig.~\ref{Fig1}. As an example, let us consider one type of memory 
devices shown in Fig.~\ref{Fig1}(B) with a magnetic fixed layer, 
whose magnetization is pinned by either an exchange bias from 
another antiferromagnetic layer or by its bulky volume, a magnetic 
free layer, whose magnetization can be changed by an external 
force such as a magnetic field, and a spacer layer of either metal 
or insulator separating two magnetic layers. No matter what is the 
source of resistance and MR in particular, the device has a higher 
and a lower resistive states, respectively, when two magnetizations 
are antiparallel or parallel to each other \cite{kundt1893hall,
Smith1929,Pugh1932}. Due to the coercivity of magnetic materials, 
when a magnetic field parallel to the magnetization of the fixed 
layer is swept up and down, a magnetic hysteresis loop is formed 
as the device moves between the two resistive states. 
This results in a parallelogram MR-curve.   
Other commonly-observed MR-curves are a butterfly-cross called 
butterfly magnetoresistance (BMR) of either upward (A) and downward 
(B) ones \cite{Ohta2021,Taniguchi2020,Mukherjee2010,Li2009} as  
shown in Fig.~\ref{Fig2}(A-B), and an MR-curve with one peak and 
one valley, called antisymmetric magnetoresistance (ASMR), as shown in 
Fig. 2(C-D) \cite{Antisymmetric2}. BMR was found in various magnetic 
materials, including van der Waals layered magnetic materials and 
strongly correlated materials, as well as many traditional magnetic 
materials \cite{Ohta2021,Taniguchi2020,Mukherjee2010,Stankiewicz2006,
Li2009,Lifan2018,Gilbert2015,Chen2018,LiP2009,Ennen2016,Mihai2008,
Liu2021,Li2017,Nguyen2011,Yujun2020,Cui-Zu2013,Huang2020,Lu2009,
Li2019,Tang2003,Miao2022FP,Klein2018,Zhang2014,Wegrowe1999,Bass1999,
Miyazaki1995,Binasch1989,Fe3O4,hirsch1959,tatsumoto1960}, at both 
high and low temperatures, in strong and weak magnetic fields, while 
people observed less common ASMR in topological Hall effect materials 
\cite{Wang2018,Zhang2021,Qin2019,Antisymmetric1}, antiferromagnetic 
topological insulators \cite{He2018}, magnetic multilayers  
\cite{Cheng2005,lin2007} and $\rm FeGeTe$ heterostructures 
\cite{Antisymmetric2}. The observation of BMR can date back to the 1950s 
\cite{hirsch1959,tatsumoto1960}. 
Although both BMR and ASMR were widely 
observed, the explanations in the literature, often involving detailed 
microscopic MR mechanisms, are different for different systems. 
For layered films such as metallic multilayers $\rm [Fe/Cr]_n$, 
$\rm [Co/Cu]_n$, and van der Waals $\rm ClI_3$ layers 
\cite{Klein2018,Bass1999,Miyazaki1995,Binasch1989}, complicated strong 
or weak electron scatterings involved magnetizations of adjacent layers 
were used to explain all kinds of MR-curves. In magnetic nanowires 
\cite{Lu2009}, FeO film \cite{Stankiewicz2006}, $\rm Co/HfO_2/Pt$ 
sandwich structures \cite{Li2017}, etc., BMR was attributed to the 
anisotropic MR effects that depend on the relative current and 
magnetization orientation. Electron-magnon scattering in systems like 
$\rm Fe_5GeTe$ van der Waal nanostructures \cite{Ohta2021}, FePt films 
and nanowires \cite{Mihai2008,Nguyen2011}, and 2D layers of $\rm Ag_2CrO_2$ 
antiferromagnetic films \cite{Taniguchi2020}, where resistance depends not 
only on magnetization but also on the applied fields and the temperature, 
is associated with BMR observation. In traditional magnets like $\rm Fe_3O_4$ 
films, the electron scattering, and tunnelling at the interfaces of nanograins 
\cite{Fe3O4} or scattering by the magnetization structures induced by fields 
and anti-phase boundaries \cite{LiP2009} were claimed to be responsible to the 
observed BMR. In many 2D materials, BMR in $\rho_{xx}$ is believed to be due 
to the quantum anomalous Hall effect (QAHE) \cite{Yujun2020,Cui-Zu2013}. 
The transverse BMR is reported in some planar Hall effects \cite{Stankiewicz2006,
Li2019,Tang2003}. In summary, both BMR and ASMR were attributed to 
very detailed microscopic interactions in the literature so far. 
The explanations lead to an impression that microscopic 
interactions are essential for these universal curves.  
People did relate the BMR to magnetization reversal and hysteresis. 
Magnetization reversal undoubtedly occurs in all magnetic materials, 
but BMR sometimes occurs, and other time does not. 
A simple universal route leading to their observation is lacking.

Here we would like to ask whether the universal BMR and ASMR have a simple 
general route independent of the origins of MR. This is a sensible question 
because most MR-curves of all magnetic materials with magnetic hysteresis, if 
not all, can be grouped into one of the above three types or their variations: 
Parallelogram-shape, BMR, and ASMR. Since the parallelogram-shaped MR-curves 
have a simple picture mentioned above, there is no reason to believe that BMR 
and ASMR would be different. 

\section{The physics of BMR and ASMR}

The resistance is a state function. For a magnetic system of a given 
magnetization distribution (magnetic/spin structure) and given external 
conditions such as the temperature, strains, external magnetic field, etc., 
the resistance is fixed. Under a given magnetic field, a system may have 
one or more than one possible stable/metastable magnetic structure. 
If a system has only one stable magnetic structure, then the MR curve, no 
matter how complicated it might be, has no hysteresis. Otherwise, the MR 
curve has hysteresis when an external magnetic field is swept up and down 
in a fixed direction. Of course, hysteresis is a general feature 
of magnetic materials due to its coercivity.  

An MR curve reflects the evolution path of the magnetic structure of a system. 
Whether an MR-curve is a parallelogram, a BMR, or an ASMR depends on 
whether the resistance of magnetic structures in a large positive magnetic 
field and in a large negative magnetic field are similar or different. 
When the resistances of large positive and negative magnetic fields are not 
too different, an MR-curve will be either a BMR or an ASMR, independent of 
the specific origin of the resistance. If the MR passes through two higher 
(lower) resistance states in sweeping-up and sweeping-down processes,
the MR-curve displays two crossed peaks (valleys) and results in
an (A) upward (downward) BMR, as shown in Fig.~\ref{Fig2}(A) or (B). 
However, if the MR passes through one higher and one lower resistance
state in sweeping-up and sweeping-down processes, respectively, 
the MR-curve displays one peak and one valley and becomes ASMR, as 
shown in Fig.~\ref{Fig2}(C) or (D). This simple picture is behind 
various magnetoresistance-related experiments on microscopic mechanisms  
although MR-curves can have different shapes from system to system. 
The coercivity field largely determines their locations of MR-loops while 
microscopic details modify their shapes, not their overall features.
\begin{figure*}
\centering
\includegraphics[width=2\columnwidth]{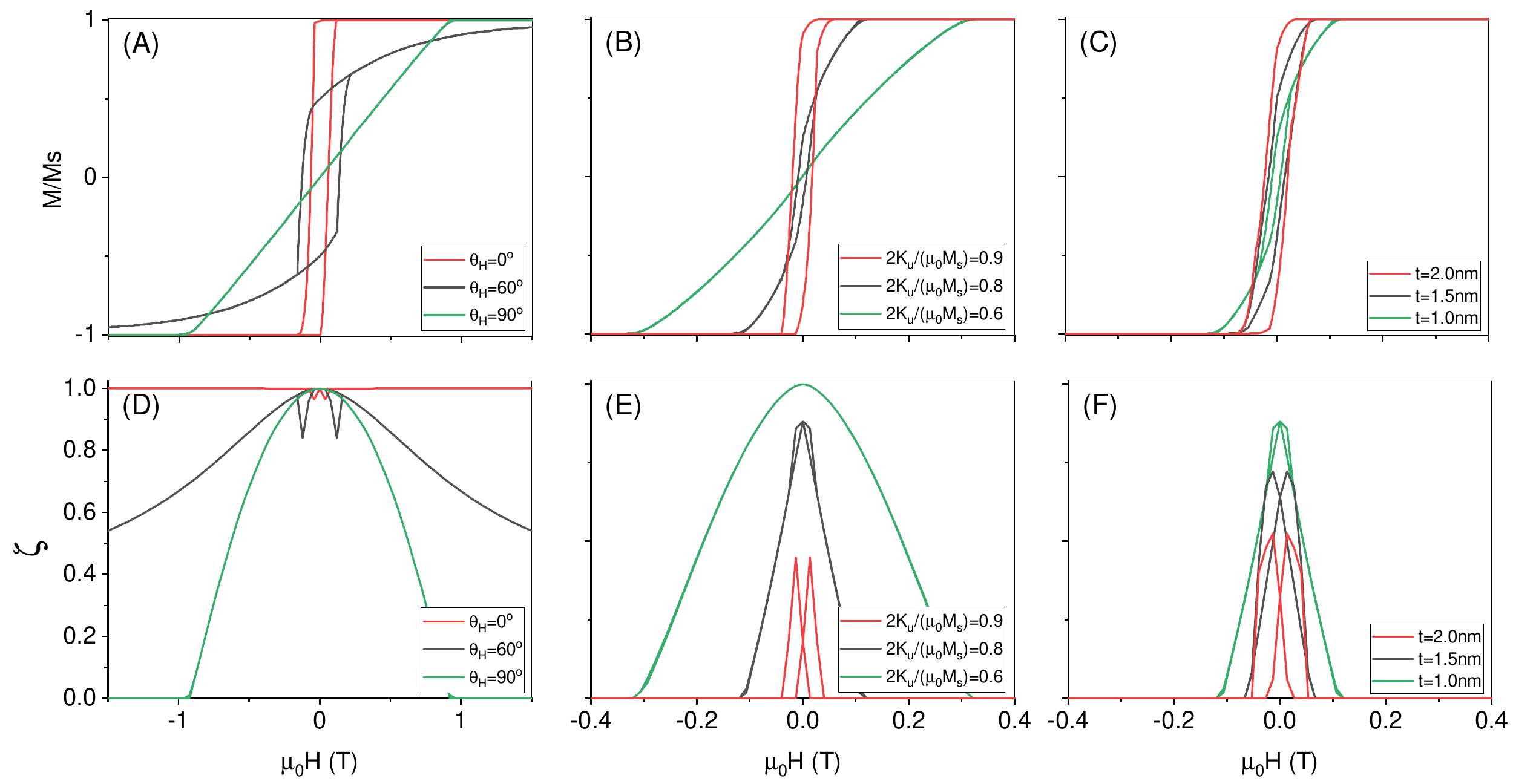}
\caption{MH-curves and MR-curves for various field directions, 
crystalline anisotropy, and sample thickness. 
(A) $m_z$ vs. $H$ along various directions. 
MH-curves display hysteresis when the field is not 
perpendicular to the easy-axis ($\hat{x}$).
(B) $m_z$ vs $H$ along $\hat{z}$ for various crystalline 
anisotropies. As $K_{\rm u}$ decreases, hysteresis disappears.
(C) $m_z$ vs $H$ along $\hat{z}$ for various sample thickness 
and for a fixed crystalline anisotropy. 
Hysteresis becomes wider as thickness grows.
(D)-F) Corresponding MR-curves of (A)-(C) with the 
resistance of $\zeta (\mathbf{m})= m_x^2$.
A BMR appears only when a hysteresis exists.  
} \label{Fig3}
\end{figure*} 

Furthermore, with the generalized Ohm's law in magnetic material, these 
principles can explain why most BMRs occur in longitudinal resistance 
measurements and are very rare in Hall resistance measurements. 
For a given magnetic material, its resistance, in general, depends on 
the magnetization when all other material parameters and their environment 
are fixed. Without losing any generality, let us define the current direction 
along the $x$-axis and transverse voltage measurement along the $y$-direction. 
The longitudinal and transverse resistance can be expressed as $R_{xx}=R_1
+A_1M_x^2$ and $R_{xy}=R_1M_z+A_1M_xM_y$, respectively according to the 
generalized Ohm's law in amorphous or polycrystalline magnetic materials 
\cite{yin,yin1,AMR2022}. $R_1$ and $A_1$ are material parameters whose values 
depends on microscopic interactions. They describe the anomalous Hall effect 
and the usual anisotropic MR (as well as the planar Hall effect), respectively. 
The above resistances are general for homogeneous systems and independent 
of electron scattering mechanisms that give rise to the resistance. 
For inhomogeneous systems, the generalized Ohm's law should refer to the 
resistivity, and magnetization and coefficients in resistances formula 
above should be properly averaged. When a magnetic field $H$ is swept 
up and down along a direction not exactly perpendicular to the magnetic 
easy-axis, the stable magnetic structures in a large positive magnetic field 
and a large negative magnetic field are two opposite magnetizations 
of $(M_{x0}, M_{y0}, M_{z0})$ and $(-M_{x0},-M_{y0}, -M_{z0})$. 
The system transforms from one state into the other 
through different paths in sweeping-up and sweeping-down processes. 
Since $R_{xx}$ is a function of $M_x^2$, no matter what $M_{x0}$ is, the 
resistances, $R_1+A_1M_{x0}^2$, in the two extreme states are the same. 
When $M_x(H)$ moves between $M_{x0}$ and $-M_{x0}$, $R_{xx}$ forms peaks 
and valleys and results in a BMR or an ASMR. Unlike $R_{xx}$ which  
depends only on $M^2_x(H)$, $R_{xy}$ depends on $M_x(H)M_y(H)$ and 
$M_z(H)$ at the same time. $R_{xy}$ at the two extreme fields takes 
different values of $2R_1M_{z0}$. This explains why most BMR appears in 
longitudinal resistance measurements but is rare in $R_{xy}$-measurements. 
However, when the magnetic field is in the $xy$-plane, i.e., the plane of 
applied current and voltage measurement. $M_{z0}$ is zero such that the 
two opposite stable magnetization states have approximately the same 
resistances. $R_{xy}$ can have peak and valley, resulting in either a 
BMR or an ASMR. This is why these two phenomena can be observed in some 
planar Hall measurements \cite{Stankiewicz2006,Tang2003,Li2019}.

\section{Demonstration of principles with toy models}

Whether a BMR or an ASMR appears depends only on whether the evolution of 
the magnetization has a hysteresis, and whether the resistances in two 
extreme states in large positive and negative magnetic fields are similar. 
In experiments, various factors can affect the appearance of BMRs,
including anisotropy, thickness, temperature, etc.
\cite{Ohta2021,Taniguchi2020,Li2009,Li2017,Klein2018}.
A BMR appears usually in a system with a strong anisotropy and at the low 
temperature. It appears sometimes in a thicker sample \cite{Li2009} and 
sometimes in a thinner one \cite{Ohta2021}. People knew that all these 
factors somehow affect the magnetization reversal through which system 
changes facilitate or prohibit the presence of a BMR, but a universal 
simple picture showing how and why a BMR occurs and does not occur in a 
specific system is lacking. Here, with the simple principles mentioned above, 
we use toy models to show these factors actually influence BMRs by changing 
the easy-axis and hysteresis, which are essential for BMRs and ASMRs. 

Our toy model is for a ferromagnetic sample whose magnetic energy is
\begin{equation}
\mathscr{E}=\int_V \{A|\nabla\mathbf{m}|^2
-K_{\rm u}m_z^2-\mu_0M_{\rm s}[\mathbf{H}+\mathbf{H}_{\rm d}]
\cdot\mathbf{m}\} \,{\rm d}\mathbf{x}^3,
\label{Etotal}
\end{equation}
where $\mathbf{m}$, $A$, $K_{\rm u}$, $\mu_0$, $M_{\rm s}$, $\mathbf{H}$ 
and $\mathbf{H}_{\rm d}$ are the magnetization unit vector, the Heisenberg 
exchange stiffness, the perpendicular magneto-crystalline anisotropy, the 
vacuum permeability, the saturation magnetization, the external magnetic 
field, and the dipolar field, respectively. If not stated otherwise, the 
sample size is $\rm 100 \times 20\times 2\,nm^3$, and model parameters 
are $A=3\,\rm pJ/m$, $M_{\rm s}=0.86\,\rm MA/m$, and $K_{\rm u}=0.3\, 
\rm MJ/m^3$, around typical values of common magnetic materials. 
We also consider a weak disorder to mimic a realistic situation. 
Random granular sample of average $10\,$nm grains are generated in films 
by Voronoi tessellation. Anisotropies of grains vary randomly by $10\%$ 
around its mean value. The non-linear Landau-Lifshitz-Gilbert (LLG) 
equation governs the spin dynamics of the model,
\begin{equation}
\begin{split}
\frac{\partial \mathbf{m}}{\partial t} =-\gamma\mathbf{m} \times 
\mathbf{H}_{\rm eff} +\alpha \mathbf{m} \times \frac{\partial \mathbf{m}}
{\partial t},
\end{split}
\label{llg}
\end{equation}
where $\gamma$, $\alpha$ and ${\bf H}_{\rm eff}$ are respectively the 
gyromagnetic ratio, Gilbert damping constant, and the effective field.
${\bf H}_{\rm eff}(\mathbf{x})=\frac{2A}{\mu_0M_{\rm s}} \nabla^2\mathbf{m}
(\mathbf{x})+\frac{2K_{\rm u}}{\mu_0M_{\rm s}}[\mathbf{m}(\mathbf{x})
\cdot\mathbf{u}]\mathbf{u}+{\bf H}+{\bf H}_{\rm d}$ includes the exchange 
field, the magneto-crystalline anisotropy field, the external magnetic 
field {\bf H}, and the dipolar field ${\bf H}_{{\rm d}}$. Eq.~\eqref{llg} 
is numerically solved by Mumax3 \cite{MuMax3}. A large $\alpha=1$
is used to speed up the search for static solutions at given fields.
The value of $\alpha$ shall not affect the principles of BMR and ASMR. 
The unit cell in the simulations is a cube of side $2\,$nm. 
The selection of the toy model is not exclusive. Other models 
could also be used to demonstrate the principles (see discussion). 
We use a dimensionless quantity $\zeta$ to mimic MR. 
$\zeta$ is a function of $m_x$, $m_y$, and $m_z$ ($\zeta=m_x^2$ for example). 
Coefficients related to detail mechanisms are neglected in this simple model. 
The MR ratio is just its normalization of MR$=(\zeta_{\rm max}-\zeta)/
\zeta_{\rm max}$. $\zeta$ can well capture the shapes of BMR, ASMR, and 
parallelogram-shaped MR-curves. To consider the contributions from all 
local magnetization-dependent resistivity, we average $\zeta(\mathbf{m})$ 
over the whole sample,
\begin{equation}
\zeta=\frac{1}{V}\int_V \zeta(\mathbf{m})\, {\rm d}V.
\label{zeta}
\end{equation}
where $V$ is the total sample volume \cite{Lu2009,Wegrowe1999,Chung2008}. 
\begin{figure}
\centering
\includegraphics[width=\columnwidth]{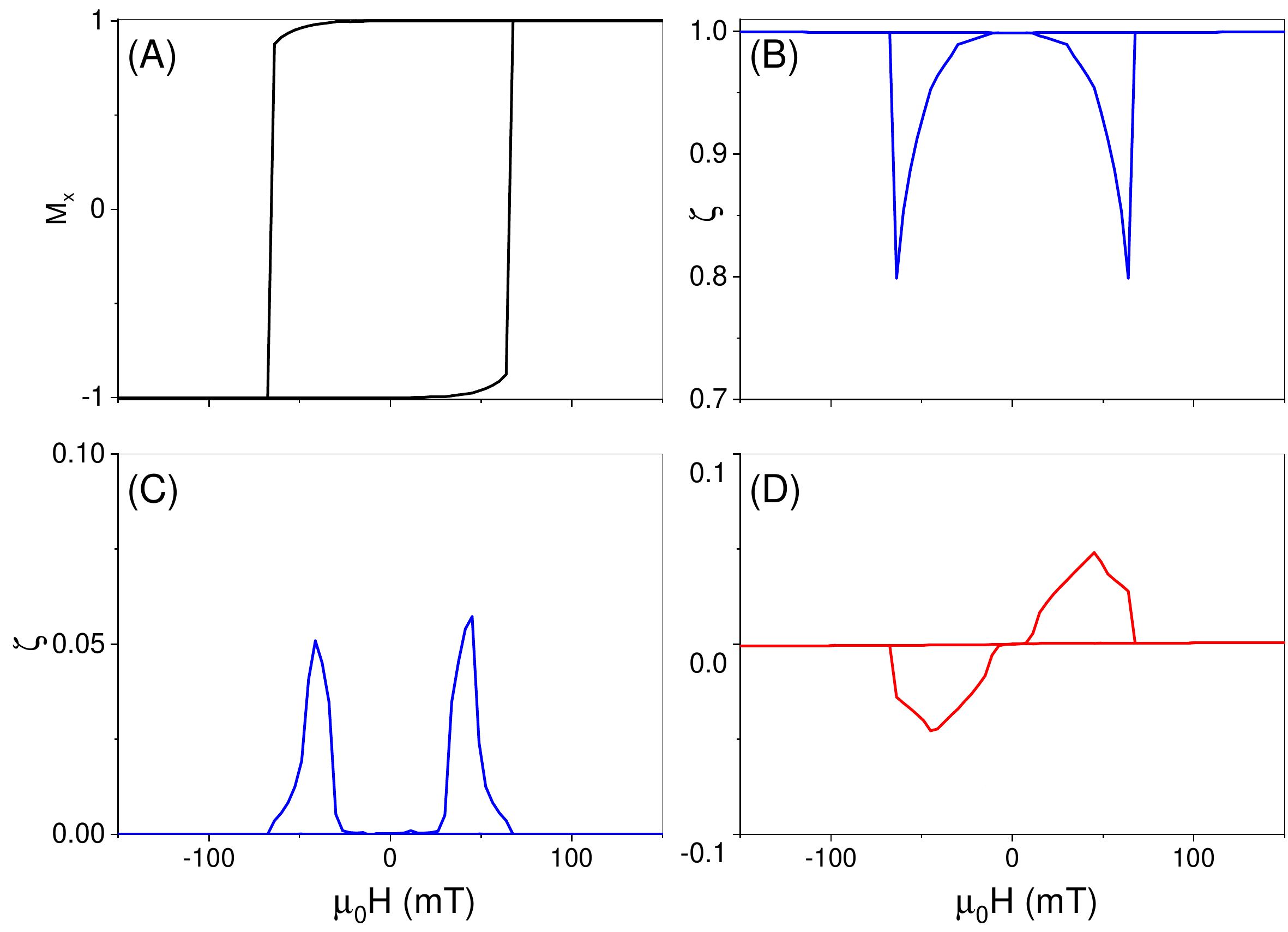}
\caption{A BMR and an ASMR model for field-sweeping in the plane. 
The fields are swept along the direction that is $0.1^\circ$ tilted 
from $\hat{x}$ to $\hat{z}$, and the current is along $x$-direction.
(A) $M_x$ as a function of $H$. A hysteresis appears with a coercivity 
field around $60\,$mT. (B) A downward BMR is obtained in the anisotropic 
magnetoresistance with two high resistance states at high fields. 
(C) An upward BMR is observed in the planar Hall resistance with two 
low resistance states at large positive and negative fields.
(D) The ASMR curve is obtained for the quantum anomalous Hall systems   
due to $M_z$ components with opposite signs in sweeping-up and 
sweeping-down processes, respectively. The resistance vanishes at 
large fields. Peaks and valleys are around the coercivity field.} 
\label{Fig4}
\end{figure}
\begin{figure}[t]
\centering
\includegraphics[width=\columnwidth]{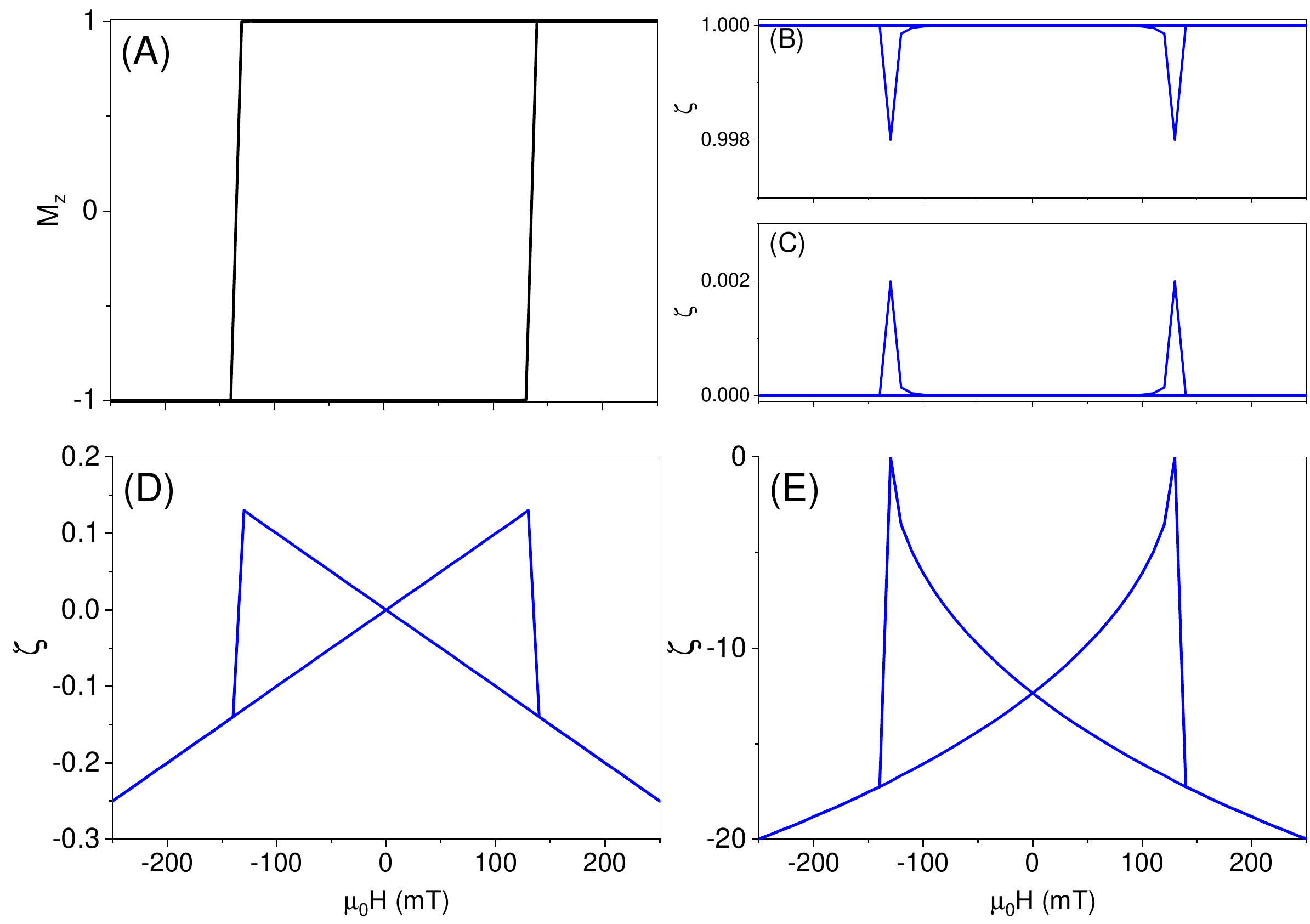}
\caption{A BMR model for a field along $0.1^\circ$ from $\hat{z}$ in the 
$xz-$plane. (A) A square-like hysteresis loop in $M_z$H-curve with a 
coercivity field around $130\,$mT. (B)-(C) Downward (B) and upward (C) BMR 
are obtained for the anisotropic magnetoresistance. The current is along 
the  $x$ and $z$ direction, respectively.
(D)-(E) The BMR from two different formulas for scatter-induced 
resistance discussed in the main text. They show minor differences 
in curvatures. Peaks and valleys are around the coercivity fields,
which is different from that of Fig.~\ref{Fig4}.} \label{Fig5}
\end{figure} 
It is known that the magnetization dynamical path relies on the angle 
between the applied field and the easy-axis \cite{stoner2005}. 
Hysteresis loops appear only when the field is not perpendicular to the 
easy-axis. Otherwise, the system has only one stable structure, and 
magnetization is reversible such that no BMRs and ASMRs are possible. 
Since a field was swept in all directions in all kinds of experiments and 
the easy-axis is sensitive to many factors such as thickness, temperature, 
etc., it is not surprising to see the appearance and disappearance of a BMR 
in similar measurements on similar samples, but with different details. 
To mimic the phenomenon, we consider a sample of $K_{\rm u}=0$ for simplicity. 
The demagnetization factors of $\hat{x}$, $\hat{y}$, and $\hat{z}$ are 
$N_x=0.86$, $N_y=0.12$, and $N_z=0.02$, and easy-axis aligns along $\hat{x}$. 
We sweep the field in different directions in the xz-plane with an angle 
$\Theta_H$ to $\hat{x}$. MH-curves for $\Theta_H=0^\circ,60^\circ,$ and 
$90^\circ$ are shown in Fig.~\ref{Fig3}(A). For $\Theta_H$ not equal to 
$90^\circ$, an MH-curve displays hysteresis loops because of the easy $x$-axis. 
For $\Theta_H=90^\circ$, or applied field perpendicular to its easy-axis, 
there is only one stable state at each given field such that the MH-curve 
is reversible and there is no hysteresis loop. The appearance of the BMR 
is closely related to that of hysteresis. We consider a resistance of $\zeta 
(\mathbf{m})= m_x^2$ for example, which can represent the anisotropic MR 
effect with a current along $\hat{x}$. The obtained curves are shown in Fig.
~\ref{Fig3}(D). When $\Theta_H\neq90^\circ$,  this model satisfies the principle 
that has the same resistance states at both large positive and negative 
fields, and the MR-curve is BMR. As $\Theta_H$ approaches $90^\circ$, it 
degenerates from BMR to a simple curve. The vanish of the BMR is due to the 
disappearance of hysteresis, regardless of specific resistance mechanisms. 
This provides a new simple picture that can explain BMR's appearance 
and disappearance when the magnetic field changes its direction in the 
experiments and simulations \cite{Stankiewicz2006,Lifan2018,Lu2009,Wegrowe1999}.

Crystalline anisotropy varies from sample to sample and leads to the 
appearance and the disappearance of the BMR and the ASMR. Perpendicular 
magnetic anisotropy can both increase and decrease \cite{Carcia1985} with 
film thickness. For materials whose perpendicular anisotropy decreases 
or even vanishes as sample thickness increases, increase of thickness may 
results in vanishing BMR in the perpendicular field-sweeping. 
For thin films of perpendicular crystalline anisotropy that is insensitive 
to film thickness, the opposite behaviour can occur: Easy-axis changes from 
perpendicular to in-plane directions as sample thickness decreases. 
With the toy model, our theory can explain disappearance of a BMR when a 
sample thickness both increases \cite{Ohta2021,Klein2018} or decreases. 
Figure~\ref{Fig3}(B) is the MH-curves for the field along the $\hat{z}$ for 
$K_{\rm u}=0.27\,\rm MJ/m^3$ [$2K_{\rm u}/(\mu_0M_{\rm s}^2)=0.6$], $0.36\,
\rm MJ/m^3$ [$2K_{\rm u}/(\mu_0M_{\rm s}^2)=0.8$], and $0.40\,\rm MJ/m^3$ 
[$2K_{\rm u}/(\mu_0M_{\rm s}^2)=0.9$] with other parameters unchanged. 
When $K_{\rm u}$ is significantly smaller than $1/2\mu_0M_s^2$, anisotropy 
is dominated by the demagnetization, and the easy-axis lies in the plane. 
There is no hysteresis since the field is perpendicular to the easy-axis. 
MR-curves, $\zeta(\mathbf{m})= m_x^2$ is shown in Fig.~\ref{Fig3}(E). 
The BMR and hysteresis disappear simultaneously. 
For another set of samples of $1\,$nm, $1.5\,$nm and $2\,$nm thick whose  
$K_{\rm u}=0.36,\rm MJ/m^3$ [$2K_{\rm u}/(\mu_0M_{\rm s}^2)=0.8$],  
MH-curves for field-sweeping along $\hat{z}$ are shown in Fig.~\ref{Fig3}(C).
Figure~\ref{Fig3}(F) is the MR-curves of $\zeta (\mathbf{m})= m_x^2$. 
With the increase of thickness, the hysteresis loop becomes fatter and the 
BMR is more pronounced because the perpendicular anisotropy is enhanced. 
The sharper BMR in a thicker film is previously attributed to the  
increase of anti-phase domain size \cite{Li2009}, very different from  
our simple universal picture.

The change of crystalline anisotropy can come from other sources. 
For example, anisotropy $K_u$ of some materials decreases with a power of 
$M(T)$ \cite{Schlickeiser}, which is sensitive to the temperature near 
the Curie temperature. The change of $K_u$ can be substantial. 
For example, the magnetic anisotropy of $1.2\,$nm CoFeB film could drop 
by 50\% as temperature increases from $300\,$K to $400\,$K \cite{Lee2017}.
As the temperature increases, perpendicular magnetic anisotropy gets 
smaller, and the easy-axis changes from out-of-plane to in-plane. 
Hysteresis, as well as BMR, thus no longer exists. In contrast to our 
universal picture of the BMR, the disappearance of the BMR at higher 
temperatures was attributed to the variation of electron scattering, 
which, in turn, was attributed to the vanish of partially disordered 
states \cite{Ohta2021,Taniguchi2020,Li2009,Li2017}.   

The resistance is a state function. BMR or ASMR curves are the 
manifestations of magnetic hysteresis. BMR and ASMR shapes and loop 
positions depend on resistance mechanisms and detailed magnetic 
properties such as coercivity fields. To further demonstrate this 
point, we use various configurations and resistance mechanisms to 
generate all kinds of BMR's and ASMR's with our toy models. 
First, we consider the in-plane fields. A smaller $K_{\rm u}=0.3\,
\rm MJ/m^3$ is used such that the easy-axis aligns with the $\hat{x}$. 
We sweep fields along the direction $0.1^\circ$ from the $\hat{x}$ in 
the $xz$-plane. The system displays a hysteresis in its MH-curve as 
shown in Fig.~\ref{Fig4}(A) with a coercivity field around $60\,$mT.
If we apply a current along the $x$-direction and consider the 
resistance of $\zeta (\mathbf{m})= m_x^2$, the MR goes down and up. 
This results in a two-valley butterfly cross illustrated in 
Fig.~\ref{Fig4}(B). BMR valleys appear around coercivity fields. 
If we consider the resistance of $\zeta(\mathbf{m})= m_xm_y$, the 
form of the planar Hall resistance. A transverse BMR can be 
obtained as shown in Fig.~\ref{Fig4}(C) that qualitatively agrees 
with the in-plane sweeping experiment \cite{Stankiewicz2006}. 
If we consider resistance in the form of $\zeta(\mathbf{m})=m_z$, 
similar to the anomalous Hall effect, a transverse ASMR is obtained, 
as shown in Fig.~\ref{Fig4}(D). Although the curve shapes are 
different, peaks and valleys appear all around coercivity fields.

To demonstrate the same BMR principles in the field sweeping along the  
perpendicular direction of a film, a larger $K_{\rm u}=4.8\,\rm MJ/m^3$ 
is used in order to maintain $\hat{z}$ as the easy-axis. 
We sweep fields along the direction that is $0.1^\circ$ tilted from 
$\hat{z}$ in the $xz$-plane. An MH-curve shows hysteresis in sweeping 
processes as shown in Fig.~\ref{Fig5}(A) with a coercivity field around 
$130\,$mT, which is sharper than Fig.~\ref{Fig4}(A). Let's still 
consider the current aligning along $x$-direction and the anisotropic 
magnetoresistance of $\zeta (\mathbf{m})= m_x^2$. The resistance goes 
up and down, resulting in an upward BMR illustrated in Fig.~\ref{Fig5}(B). 
If the current is perpendicular to the film, the magnetoresistance 
becomes $\zeta (\mathbf{m})= m_z^2$. the MR goes down and up, resulting 
in two valleys in the MR-curves, i.e., a downward BMR shown in Fig.
~\ref{Fig5}(C), which qualitatively agrees with experiments \cite{Li2017}. 
In this configuration, peaks and valleys appear at the coercivity fields, 
which differ from those of Fig.~\ref{Fig4}. One can also reproduce 
BMR with other microscopic mechanisms, such as magnon-electron scattering, 
where the external fields tune the magnetization-dependent resistance 
\cite{Taniguchi2020,Mihai2008,Nguyen2011}. If the resistance is linearly 
in the field, i.e. $\zeta(\mathbf{m})=m_z\mu_0H$, a BMR curve similar to  
that in Ref.~\cite{Mihai2008,Nguyen2011} can be reproduced as shown in 
Fig.~\ref{Fig5}(D). If we consider the MR in the form of 
$\zeta(\mathbf{m})=-m_z\sqrt{\Delta /(\Delta +2)}$, where $\Delta 
\propto H-H_{\rm a}$ measures the difference between the applied field 
and the anisotropy field, a BMR of Fig.~\ref{Fig6}(E) can be obtained,
where we choose $\Delta= \mu_0(H-H_{\rm a})$ as a demonstration.
This result qualitatively agrees with experiments \cite{Taniguchi2020}. 

\section{discussions and conclusion}

We used toy models governed by the LLG equation to demonstrate BMR and 
ASMR in various systems, because it is compatible with experiments involving 
incoherent magnetization reversal such as those in Refs. \cite{Mihai2008,
Nguyen2011,Lu2009}, Other models could also be used in different scenarios. 
For example, BMR has also been observed in systems described by 
coherent-rotational models \cite{zhao2007,zhao2019}, and other 
special reversal process such as domain wall nucleation 
and motion described by the Kondorsky model \cite{Lu2009}. 
Between two states at large positive and negative fields, 
there is a hysteresis in these systems, also consistent with the general 
picture here. If one uses the similar procedure as that in the third section, 
BMR or ASMR can also appear as long as the resistances is a function of 
magnetization, and the resistance of two states at high fields are not 
too different, see Supplementary Information for details.
\begin{figure}
\centering
\includegraphics[width=\columnwidth]{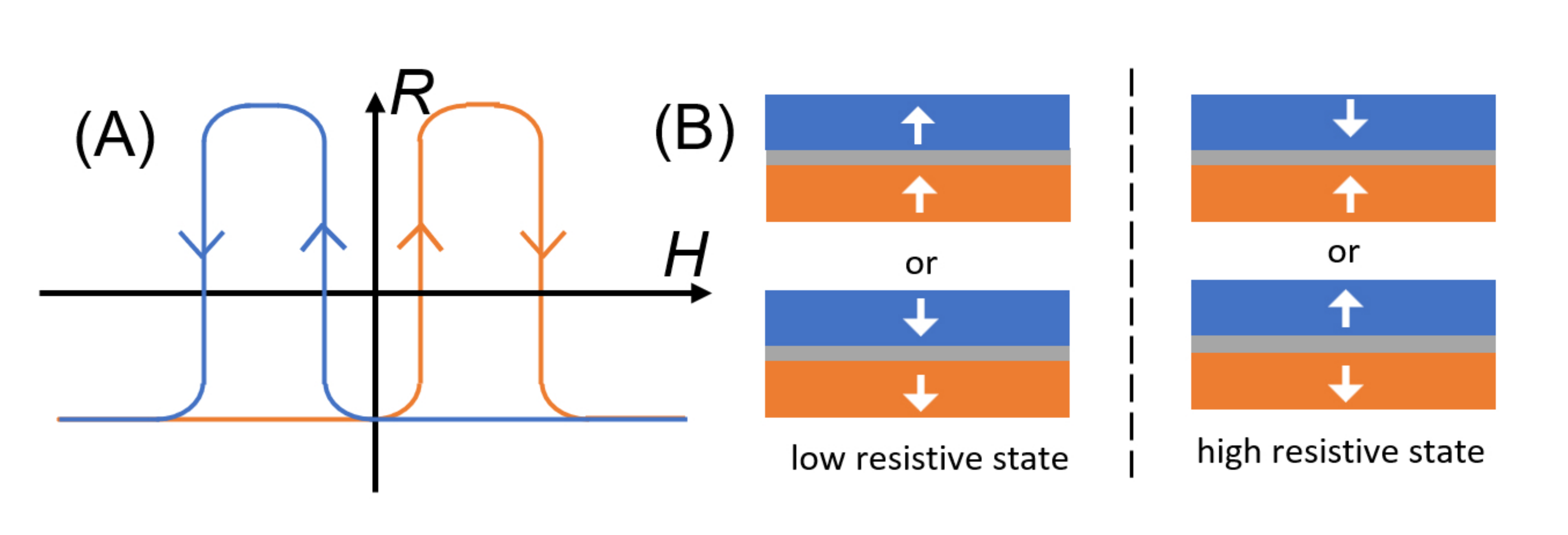}
\caption{(A) Schematics of BMR in a bilayer.
The system moves between two similar low resistive states
by passing through a high resistive state during the field 
sweeping. (B) Illustration of resistive states. 
The device is in a higher (lower) resistive state when the magnetization 
of two layers are parallel (antiparallel) with each other.}
\label{Fig6}
\end{figure}

The rules of BMR and ASMR revealed by the toy models are general.
For example, in a sample of two free layers, in contrast to a memory cell 
where only one is free, as sketched in Fig.~\ref{Fig1}(B), the system 
has a lower resistive state of two magnetizations parallel to each other 
and a higher resistive state of two antiparallel magnetization. 
When the magnetic field is swept along the $z$-direction, the system 
moves between two stable low resistive states of magnetizations 
along $\hat{z}$ and $-\hat{z}$. The magnetization of two layers 
is no longer parallel to each other during the magnetization reversals, 
and the resistance forms two peaks and an upward BMR \cite{Klein2018,
Bass1999,Miyazaki1995,Binasch1989} as shown in Fig.~\ref{Fig6}. 
We also consider a system with the anomalous quantum Hall effect. 
Assume that the magnetic field is swept along $z$-direction, and the 
system switches between the ferromagnetic states of $m_z=1$ and $m_z=-1$, 
both of which have $R_{xx}$ equal to 0. However, in sweeping processes, 
the presence of magnetic structure leads to a finite $R_{xx}$, 
resulting in two peaks and an upward BMR \cite{Yujun2020,Cui-Zu2013}. 
$\rm Fe_3O_4$ films have the same resistance states at both high positive 
and high negative fields that decrease with a field strength in the same 
slope \cite{LiP2009}. Consequently, it displays a BMR. In some chiral 
magnetic materials, the topological Hall resistance is related to the 
topological charge. The system reverses between the states of $m_z=1$ and 
$m_z=-1$, which have zero topological charge and zero topological Hall 
resistances. In the sweeping-up and sweeping-down process, topological 
charges with opposite signs are generated in the system that produces one 
peak and one valley on the MR-curve manifesting an ASMR 
\cite{Wang2018,Zhang2021,Qin2019}. 

It may be important to emphasize that the exact shapes and locations in a BMR 
and ASMR are not our concerns here. The hysteresis loop of BMR and ASMR could 
be very irregular in different systems. Their general features do not rely on 
any symmetries, as shown in the toy models of Figs.~\ref{Fig4} and \ref{Fig5}. 
When the resistance mechanism has inversion symmetry, the system is more 
likely to display ASMR as the toy model shown in Fig.~\ref{Fig4}(D). 
In general, any curve can always be decomposed into symmetric and antisymmetric 
components, as was usually done in experiments. Nevertheless, this kind of 
decomposition is not meaningful unless one can attribute each of them to a 
specific source. This is, of course, an interesting question but not the aim 
of this paper. 

ASMR is less common than BMR in both $R_{xx}$ and $R_{xy}$. In many 
experiments, the magnetic field is swept perpendicular to the currents. 
For the former, $M_x(H)$ reverses between $\pm\{M_{x}\}_{\rm max}$. 
These two opposite magnetizations have the largest $R_{xx}$, 
so in sweeping-up and sweeping-down processes, $R_{xx}$ can only decrease 
first and then increase and form two valleys and a downward BMR. 
For the latter, $M_x(H)$ changes from 0 to 0 through a path. 
The initial and final magnetizations have the lowest $R_{xx}$. So whether 
in the sweeping-up or sweeping-down process, $R_{xx}$ can only rise and then 
fall, which is manifested as an upward BMR. For magnetic materials that obey 
the generalized Ohm's law, ASMR that requires one peak and one valley is 
unlikely to occur in $R_{xx}$ for either of the two common experimental settings.

In conclusion, similar to parallelogram MR, BMR and ASMR are universal MR behavior 
independent of the resistance origins. They appear as long as a system exhibits 
a hysteresis loop under sweeping-up and sweeping-down of a magnetic field, and 
its MR is approximately the same when the magnetization direction is reversed. 
From the generalized Ohm's law in magnetic materials, BMR should be very common 
in longitudinal resistance and could als occur in transverse resistance when 
the magnetic field is in the plane of applied current and voltage measurement. 
The coercivity fields and microscopic details are encoded in the positions and 
shapes of a BMR and ASMR. Although there is no inconsistent with the universal 
BMR/ASMR theory presented here, the explanations of many BMR and ASMR in 
literature \cite{Ohta2021,Taniguchi2020,Li2009,Antisymmetric2,
Stankiewicz2006,Li2017,Klein2018,Wegrowe1999,Bass1999,Miyazaki1995,
Binasch1989} are not the same as ours. 

\begin{acknowledgments}
This work is supported by the Ministry of Science and Technology through grant 
2020YFA0309600, the NSFC Grants (12074301, 11974296), and Hong Kong RGC Grants 
(No. 16300522, 16302321, 16301518, and 16301619). 
\end{acknowledgments}


%

\end{document}